\documentclass[12pt]{iopart}
\usepackage[dvips]{graphicx}
\usepackage{iopams,setstack} 
\usepackage{amssymb} 
\begin{document} 
%----------------------------------------- 
\title{Oscillations bounded by the simple pendulum and the oscillating rigid rod} 
\author{J  C Zamora, F Fajardo and J-Alexis Rodr\'iguez}
\address{Departamento de F\' isica, Universidad Nacional de Colombia, Bogota, Colombia}
\ead{jarodriguezl@unal.edu.co}

%\date{}
%\maketitle 
\begin{abstract}
The oscillation periods bounded by a simple pendulum and  an oscillating rigid rod  are illustrated using a multiple pendulum.  Oscillation periods between these two limits are obtained. A theoretical approach using the Lagrangian formalism  and the set up of three simple experiments are presented in order to probe our approach. 
\end{abstract}
%\pacs{}

\maketitle
%------------------------------------------------------------- 
 
\section{Introduction} 
In any introductory physics course the simple pendulum and the oscillating rigid rod are usual topics of study and they are not only discussed theoretically but moreover are part of the basic former experiments done by students. It is because they are simple and are used to illustrate the extension of a specific set of data to a general rule, and also they become one of the first approaches to the scientific thinking. Once the students have studied the simple pendulum and the oscillating rigid rod, then it is possible to ask them if there are other possible physical systems, such that the oscillating periods were bounded by the simple pendulum and the oscillating rigid rod. It is meaning, pendulums with an oscillating period in the shaded area of figure 1. In order to answer that question,  the students will notice that for a simple pendulum the mass is considered a point mass at the end of a string and in the case of the oscillation rigid rod the mass is distributed uniformly overall the rigid rod. Then we can propose to the students an special case of the multiple pendulum like is shown in figure 2. Equal masses  are uniformly spaced at distances $a=L/n$, with $L$  the total length of the pendulum and $n$ the number of masses in this multiple pendulum.  The case $n=1$ is going to be the simple pendulum and making the limit $n \to \infty$  it would be the oscillating rigid rod. To be clear, it is worth to mention that the whole set of masses on the string should oscillate at the same time, it is with the same oscillating period. If it was not the case then the normal modes of a string motion type in the continuum limit will be arise.

The multiple pendulum has been treated theoretically in many basic books, for instance \cite{fisica, clasica}  and a detailed analysis has been done by M. Braun \cite{braun}. On the other hand some experimental approaches has been done using the paper-clip chain \cite{clips} and this problem was figured out in a theoretical way in reference \cite{ingenieros}. In section 2 some theory about the multiple pendulum is reviewed. The equations of motion of a multiple pendulum are derived and expressions for the oscillating period are obtained. The obtained mathematical expressions agree with the previous ones cited \cite{braun,clips,ingenieros}. The experimental analysis is presented in section 3 and our conclusions are in section 4.

\section{Some theory}

Consider a pendulum of total length $L$ with $n$ masses spaced $a$ in such way that they are coupled as it is shown in figure 2. In general each pendulum have an angle $\phi_k$, and therefore the kinetic and potential energy are in each case, 
\begin{eqnarray}
T_k = \frac 12 m_k \sum_{i=1}^k \sum_{j=1}^k a_i a_j \dot \phi_i \dot \phi_j \cos(\phi_i-\phi_j) ,\nonumber \\
V_k = m_k g \sum_{i=1}^k a_i (1-\cos\phi) ,
\end{eqnarray}
and then the Lagrangian is
\begin{equation}
{\cal L}=\sum_k^n T_k-V_k  .
\end{equation}
The case under analysis corresponds to take $a_i=a_j=a$, $m_k=m$ and the small oscillation approach  $\cos (\phi_i-\phi_j)\sim 1$ and $\cos \phi_i\sim 1-\phi^2_i/2$. After some algebraic manipulation the Lagrangian can be written as,
\begin{equation}
{\cal L}=\frac a 2 \sum_{k=1}^{n}\left[ m_k \left( \sum_{i=1}^k \left( a \dot \phi_i^2-g \phi_i^2 + 2a\sum_{i\neq j}^k \dot \phi_i \dot \phi_j \right) \right)   \right] ,
\end{equation}
and then the Euler-Lagrange equations are
\begin{equation}
\sum_{k=1}^n \left[m_k \left( \ddot \phi_i+ \frac g a \phi_i +\sum_{i\neq j}^k \ddot \phi_j \right)  \right]=0  .
\end{equation}
with $i \leq k \leq n$. 

Now, small oscillations approach is used,
\begin{equation}
\phi_1 \sim \phi_2 \sim \phi_3 \sim \dots \cong A \cos(\omega t + \theta) .
\label{small}
\end{equation}
Notice that it is implying that the coupled masses are oscillating with the same amplitude and at the same frequency. It is because an oscillating rigid rod should be obtained when the number of masses in the multiple pendulum is tending to infinity. Then, expression (4) is a set of coupled differential equations that due to the symmetry can be rewritten as
\begin{equation}
-\omega^2 
\left(
\begin{array}{cccc}
n & n-1 & \cdots & 1\\
n-1 & \cdots  & \cdots & 1 \\ 
\vdots & \vdots & \ddots & \vdots \\
1 & 1 & \cdots & 1 
\end{array} 
\right)
\vec \phi
+ \frac g a 
\left(
\begin{array}{cccc}
n & 0 & \cdots \\
0 & n-1 & \cdots & 0 \\
\vdots & \vdots & \ddots & \vdots \\
0 & 0 & \cdots & 1  
\end{array} 
\right)
\vec \phi
=0
\end{equation}
where $\vec \phi$ is a vector whose components are  $\phi_1,\phi_2,\phi_3 \cdots$.  The $n$ conditions must be satisfied in order to get the frequency $\omega^2$, one way is summing over all them, obtaining
\[
\omega^2 = \frac{\lambda_1}{\lambda_2} \frac{g}{a}
\]
where $\lambda_1$ is the trace of the diagonal matrix and $\lambda_2$ is the sum of the diagonals of the another matrix, they are
\begin{eqnarray}
\lambda_1 = \sum_{i=1}^n i = \frac{n(n+1)}{2} , \\
\lambda_2 = \frac{n(n+1)}{2} + 2 \sum_{k=1}^{n-1} \frac{k(k+1)}{2} .
\end{eqnarray}
Using the above expressions and taking into account that $a=L/n$, the oscillating period is
\begin{eqnarray}
\fl T = 2 \pi \sqrt{ \frac{L \lambda_2}{n g \lambda_1}} \nonumber\\
 = 2 \pi \sqrt{\frac{L}{n g}} \sqrt{\frac{2}{n(n+1)} \left( \sum_k^n k^2-\sum_k^n k \right)-1} \nonumber \\
 = 2 \pi \sqrt{\frac 2 3 \frac{L}{g}} \left( 1+\frac{1}{2n} \right)^{\frac 12} \label{periodo}
\end{eqnarray}
and finally it is,
\begin{equation}
T=2 \pi \sqrt{\frac 2 3 \frac{L}{ng}}\left( n+\frac 12 \right)^{\frac 12} .
\label{periodo2}
\end{equation}
which agrees with similar expressions found in references \cite{braun,clips,ingenieros}. Notice that the equation (\ref{periodo}) in the limit $n=1$ is reduced to the standard form of the oscillating period for a simple pendulum and in the limit of $n \to \infty$, it is reduced to the oscillating period of a rigid rod of total length $L$,
\[
T_{rod}= 2 \pi \sqrt{\frac{2}{3} \frac L g} .
\]
Therefore the expression (\ref{periodo}) is useful to study the multiple pendulum as pendulums with oscillating periods in the bordered area of figure 1.

\section{Experimental Set up}
 Three different experimental arrangements are presented.
The first goal is to observe the behavior of the oscillation period of the multiple pendulum as a function of its linear density mass. In the second experimental set up,  the linear density mass is fixed and the total length of the multiple pendulum is changed. And finally, it is experimentally shown that the oscillating period of a multiple pendulum under some conditions can be between the oscillating period of a simple pendulum and the oscillating period of a rigid rod. The materials used to do these experiments are thread segments, a photogate time PASCO ME-9215A and 36 masses. The masses were rubber balls of a radio of 2.55 cm and  8.4 gr. Although the oscillating period of a pendulum in principle is independent of the attached mass, the masses were carefully chosen taking into account a difference of about $\pm 0.2$ gr in their weight. Using a common needle through the balls, they were attached to the thread. First of all, it is necessary to verify that the whole set of masses are oscillating with the same period and the same $\phi_i$ angle, it means if equation (\ref{small}) is a good approach. To verify that condition, the oscillating period for each attached mass in the multiple pendulum was measured, as is shown in figure 3a. For a pendulum of eight attached massses, the masses 4th, 5th, 6th and 7th presented a  maximum deviation in the oscillating period of about $0.2$ $\%$ respect to the 8th mass. Where the masses are counted beginning with the mass nearest to the fixed point. Similar results are obtained for a pendulum of four attached masses and for a pendulum of two attached masses.

\subsection{Experimental Set up 1}
The total length of the multiple pendulum $L=1.12$ m was fixed and the masses were equally spaced over the total length $L$, see figure 3a. The number of masses used were from 1 to 32 units.  The oscillating period was measured for ten oscillations in each case and the results are shown in table 1. In table 1, the theoretical prediction ($T_{th}$) using equation (\ref{periodo}) is compared to the experimental value obtained ($T_{exp}$) and finally the deviation between these two values is shown. Notice that the error is not bigger than $1.1 \%$. In figure 4, the oscillating period versus the number of masses $n$ have been plotted and  the dominance of the factor $1/\sqrt{n}$ from equation (\ref{periodo}) is clear. Furthermore from figure 4, increasing the number of masses, the curve is going asintotically to the value $T=1.737$ which is the value of the coefficient for an oscillating rigid rod. Experimentally, the upper limit is $n=\Vert L/d \Vert$  which is the minimum integer function of the rate between the total length $L$ and the diameter of the masses $d$. For this case, the limit would be $44$ masses, and using equation (\ref{periodo}), the period is $1.747$ \textit{s}, which is deviated $0.6$ $\%$ from the oscillating period of the rigid rod. In figure 5 the square of the oscillating period versus the inverse of $n$ have been plotted.  The linear behavior is established.

\begin{table}[h]
\caption{The oscillating periods changing the number of masses $n$ and $L=1.12$ m. The $\epsilon$ $(\%)$ is the deviation of the experimental value respect to the theoretical prediction using equation(\ref{periodo}).}
\begin{indented}
\item[]
%\begin{center}
\begin{tabular}[t]{@{}llll}
\br
$n$ & $T_{exp} \pm 0.002 (s)$ & $T_{th}(s)$ & $\epsilon (\%)$ \\ \mr
1 & $2.104$ & 2.127 & 1.1  \\ \mr
2 & $1.940$ & 1.942 & 0.1  \\ \mr
4 & $1.845$ & 1.842 & 0.2  \\ \mr
8 & $1.798$ & 1.790 & 0.4  \\ \mr
16 & $1.766$ & 1.764 & 0.1 \\ \mr
32 & $1.757$ & 1.751 & 0.3 \\ \br
\end{tabular}
\end{indented}
%\end{center}
\end{table}

\subsection{Experimental Set up 2}

Following the idea of equation (\ref{periodo}), an experiment is set up in order to test variations to the number of attached masses. To do it, the linear density mass $n/L$ is fixed and it is done chosen the number of masses $n=10$ by each meter of total length. The total length of the multiple pendulum was cut according to $(n-i)/n$ where $i \in Z$ with $0 \leq i \leq n$,  it is illustrated in figure 3b. Again the oscillating period for ten oscillations was measured in each case.  Figure 6 has the results for this experiment. 
%\begin{table}
%\begin{center}
%\begin{tabular}[t]{||c|c||}
%\hline
%$n$ & $T(s)$ \\ \hline
%2 & $0.819 \pm 0.001$ \\ \hline
%3 & $0.967 \pm 0.002$ \\ \hline
%4 & $1.106 \pm 0.001$ \\ \hline
%5 & $1.216 \pm 0.001$ \\ \hline
%6 & $1.324 \pm 0.001$ \\ \hline
%7 & $1.425 \pm 0.002$ \\ \hline
%8 & $1.523 \pm 0.002$ \\ \hline
%9 & $1.604 \pm 0.003$ \\ \hline
%10 & $1.695 \pm 0.003$ \\ \hline
%\end{tabular}
%\end{center}
%\caption{Data fixing the linear density mass $n/L$ }
%\end{table}
The oscillating period ($T(s)$) after a power regression on data of figure 6 is
\begin{equation}
T= (0.511 \pm 0.037) (n+(0.524 \pm 0.016))^{(0.509\pm 0.003)} .
\end{equation}
It is worth to compare it to equation (\ref{periodo2}), which is $T=0.519 \sqrt{(n+0.5)}$. The correlation coefficient is 0.994 and comparing the experimental fit to the theoretical expression, the coefficient is deviated $1.5$ $\%$ and the power number $1.8$ $\%$. The number inside the parenthesis, 0.524, is deviated about $5$ $\%$, it is large due to extra contributions of $n$   \cite{braun}.

\subsection{Experimental Set up 3} 

Finally, the number of atacched masses (1, 2, 4 and 8) equally spaced is fixed but the total length $L$ is changed.  The results are in figure 7. Equation (\ref{periodo}) has been used to draw in  figure 7 the different oscillating periods between the simple pendulum and the limit case of a rigid rod. Notice how increasing the number of masses  equally spaced over a total length $L$,  the case of the oscillating rigid rod is gotten. Therefore the period of a multiple pendulum is between  these two limits: a simple pendulum and the limit case of an oscillating  rigid rod.  The different experimental set up has shown how the equation (\ref{periodo}) is a good expression for the oscillating period of a multiple pendulum. 

%\begin{table}
%\begin{center}
%\begin{tabular}[t]{||c|c|c|c|c||}
%\hline
%$L(m \pm 0.001)$ & $T_1(s)$ & $T_2(s)$ & $T_4(s)$ & $T_8(s)$ \\ \hline
%1.120 & $2.099 \pm 0.003$ & $1.940 \pm 0.002$ & $1.845 \pm 0.002$ & $1.798 \pm 0.002$ \\ \hline
%1.000 & $1.993 \pm 0.004$ & $1.831 \pm 0.002$ & $1.753 \pm 0.002$ & $1.706 \pm 0.001$ \\ \hline
%0.800 & $1.790 \pm 0.009$ & $1.643 \pm 0.002$ & $1.563 \pm 0.002$ & $1.529 \pm 0.001$ \\ \hline
%0.600 & $1.546 \pm 0.003$ & $1.422 \pm 0.003$ & $1.352 \pm 0.002$ & $1.315 \pm 0.002$ \\ \hline
%0.400 & $1.281 \pm 0.004$ & $1.166 \pm 0.001$ & $1.106 \pm 0.001$ & $1.076 \pm 0.001$ \\ \hline
%0.200 & $0.927 \pm 0.003$ & $0.819 \pm 0.001$ & $0.781 \pm 0.001$ & $\cdots$ \\ \hline
%\end{tabular}
%\end{center}
%\caption{The oscillating period $T_n(s)$ for $n$ equally spaced masses over the total length $L$.}
%\end{table}

\section{Conclusions} 

A particular case of the multiple pendulum in the small oscillations approach has been treated theoretically and experimentally. Theoretically, an expression for the oscillating period is obtained, it is equation (\ref{periodo}). It is a function of the total length $L$ of the pendulum and the number of attached masses $n$. The expression (\ref{periodo}) is reduced to the standard oscillating period of a simple pendulum when $n=1$ and the oscillating period of a rigid rod is gotten when $n \to \infty$. Three basic experiments have been described in order to show the validity of the theoretical approach. Oscillating periods between a simple pendulum and a rigid rod have been gotten changing the number of masses in a multiple pendulum. When the linear density mass of the multiple pendulum is increased, the oscillating period is going to the limit case of a rigid rod while if it is decreased the period is tending to the one of a simple pendulum.

\newpage

\begin{figure}
\begin{center}
\includegraphics[width=8cm, height=8cm]{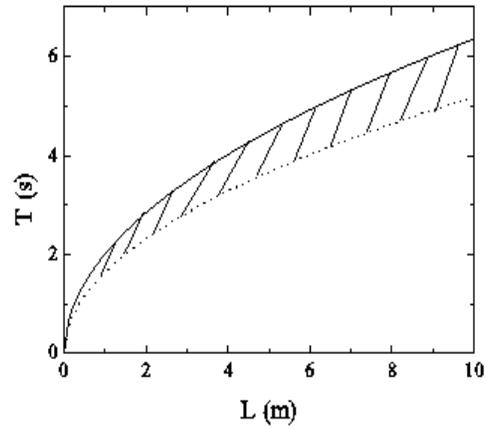}
\end{center}
\caption{The oscillating period versus the total length, the bordered area is between the oscillating period of the simple pendulum (solid line) and the oscillating period of a rigid rod (dotted line).} 
\end{figure}

\begin{figure}
\begin{center}
\includegraphics[scale=1]{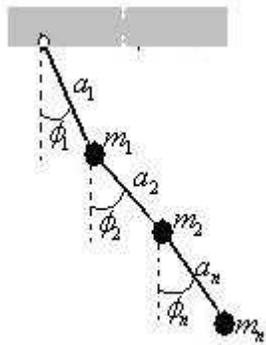}
\end{center}
\caption{The general multiple pendulum with $m_i$ masses doing an angle $\phi_i$ and  length $a_i$. A special case is $a_i=a$,  $\sum a_i=L$, $m_i=m$ and the whole set of atacched masses doing the same angle respect to fixed point.}
\end{figure}

%\newpage

\begin{figure}
\begin{center}
\includegraphics[scale=1]{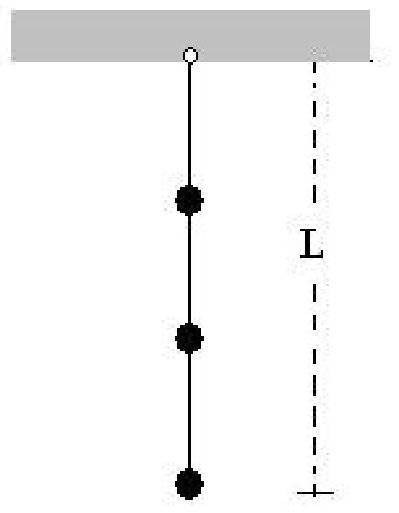} \quad
\includegraphics[scale=1]{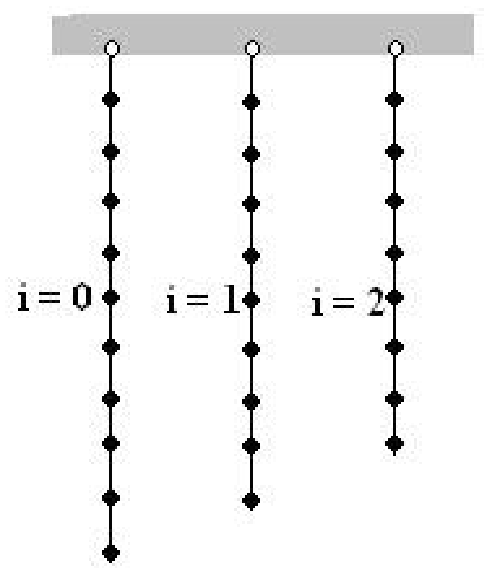}
\end{center}
\caption{Schematic view of the multiple pendulum studied. The masses $m$ are equal and they are spaced over the total length $L$. Figure 3a on left side corresponds to the simplest one set up. Figure 3b illustrates when the total length is changed and the linear density mass is fixed.}
\end{figure}

\begin{figure}
\begin{center} 
\includegraphics[width=8cm, height=8cm]{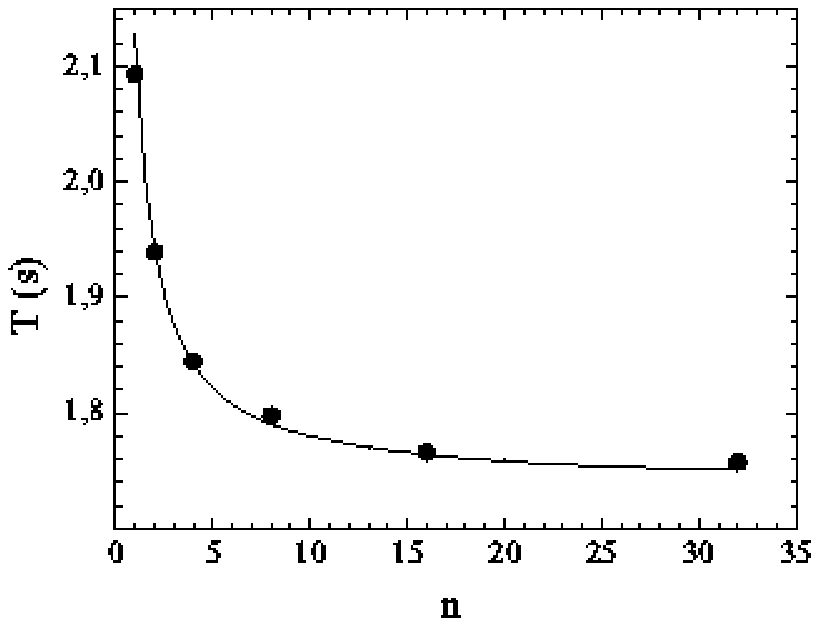}
\end{center}
\caption{The oscillating period $T(s)$ versus the number of masses $n$. The dots are the experimental data and the solid line corresponds to the theoretical prediction.} 
\end{figure} 

%\newpage

\begin{figure}
\begin{center} 
\includegraphics[width=8cm, height=8cm]{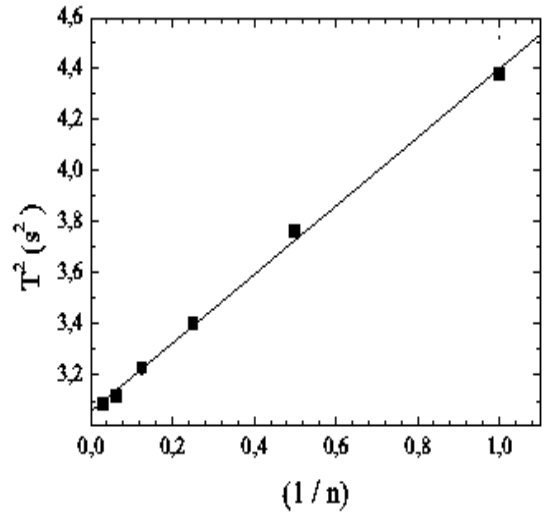}
\end{center}
\caption{The oscillating period square $T(s)$ versus the inverse of the number of masses $1/n$. The dots are the experimental data and the solid line corresponds to the linear fit.} 
\end{figure}

%\begin{figure}
%\begin{center}
%\includegraphics[angle=0]{figur6.ps}
%\end{center}
%\caption{An illustration of the experimental set up 2}
%\end{figure}

%\newpage

\begin{figure}
\begin{center} 
\includegraphics[width=8cm, height=8cm]{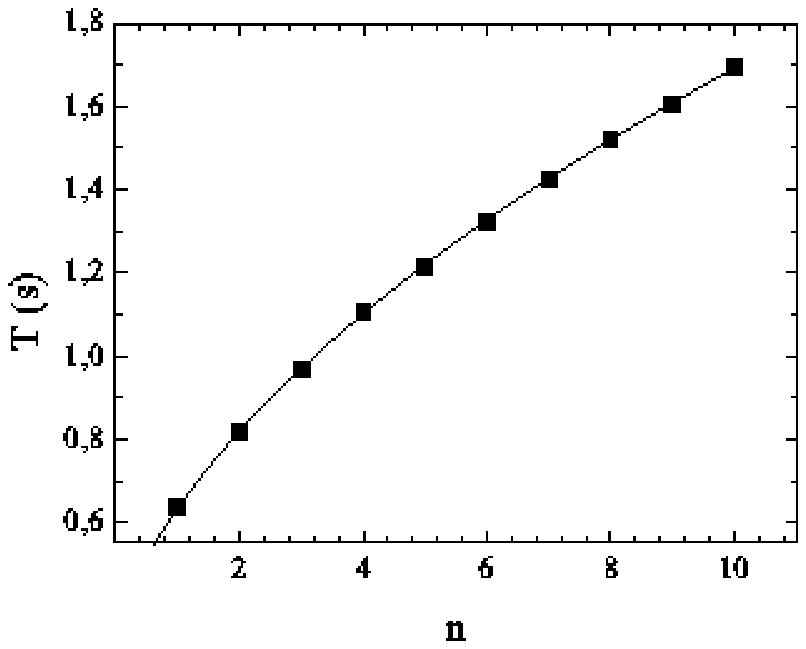}
\end{center}
\caption{The oscillating period $T(s) \pm 0.002$ versus the number of masses $n$. The dots are the experimental data and the solid line corresponds to the theoretical prediction.} 
\end{figure}

%\begin{figure}
%\begin{center} 
%\includegraphics[angle=0]{figur8.ps}
%\end{center}
%\caption{Experimental set up 3} 
%\end{figure}

%\newpage

\begin{figure}
\begin{center} 
\includegraphics[scale=1]{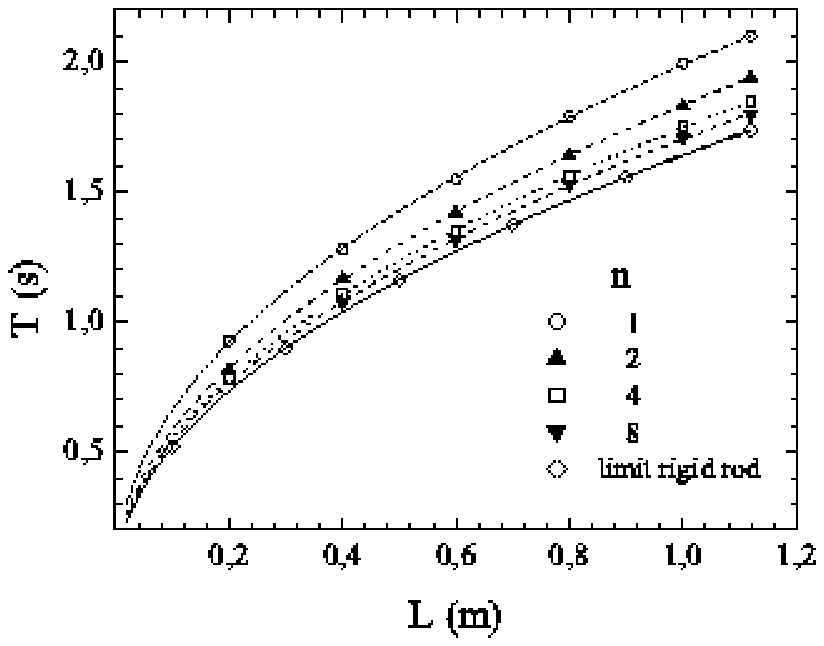}
\end{center}
\caption{The oscillating period ($T(s) \pm 0.002$) versus the total length $L(m) \pm 0.001$. The theoretical predictions for the multiple pendulum has been calculated using equation (\ref{periodo}). $n$ is the number of attached masses for each length.} 
\end{figure}


\section*{References} 
\begin{thebibliography}{99}
 
\bibitem{fisica} P. Fishbane, S. Grasiorowicz ans S. Thomton, {\it Physics,} Ed. Prentice Hall, 1996.

\bibitem{clasica} H. Goldstein and C. Poole, {\it Classical Mechanics}, Ed. Addison Wesley, Third Edition.

\bibitem{braun} M. Braun, Applied Mechanics {\bf 72} (2003) 899.

\bibitem{clips} D. Oliver, The Physics Teacher, {\bf 34} (1996) 446.

\bibitem{ingenieros} C. R. Wylie and L. C. Barret, {\it Advanced Engineering Mathematics}, Ed. MaGraw-Hill, 6th edition, NY 1995.


\end{thebibliography}
\end{document}